\documentclass[11pt]{article}

\pdfoutput=1

\usepackage[sort&compress,numbers]{natbib}

\usepackage{times}
\usepackage{amsmath}

\usepackage{xcolor}
\usepackage{enumitem}

\usepackage{epsf}
\usepackage{epsfig}
\usepackage{graphicx}
\usepackage{bm}
\usepackage{listings}
\usepackage{multicol}
\usepackage{wrapfig}
\usepackage{subfigure}

\lstdefinelanguage{pseudo}{
morekeywords={procedure,if,then,else,while,for,true,false,break,repeat,until,and,or,type,record,enum,return,end,null,void,int,bool,atomic,not,new},
sensitive=true,
morecomment=[l]{\#},
morecomment=[l]{//},
morestring=[b]',
morestring=[b]",
}
  
\usepackage{afterpage}
\usepackage{url}
\begin{document}

\newcommand{\superscript}[1]{\ensuremath{^{\textrm{#1}}}}
\newcommand{\subscript}[1]{\ensuremath{_{\textrm{#1}}}}

\newcommand{\TM}{\texttrademark}
\newcommand{\REG}{\superscript{\mbox{\tiny\textregistered}}}

\newcommand{\todo}[1]{\textbf{[TODO: #1]}}

\newcommand{\opfont}[1]{{\tt #1}}
\newcommand{\Collect}{\opfont{Collect}}
\newcommand{\Store}{\opfont{Update}}
\newcommand{\Register}{\opfont{Register}}
\newcommand{\Deregister}{\opfont{DeRegister}}

\newcommand{\CollectOp}{\opfont{Collect}}
\newcommand{\RegisterOp}{\opfont{Register}}
\newcommand{\DeregisterOp}{\opfont{DeRegister}}
\newcommand{\DeOrRegisterOp}{\opfont{(De)Register}}

\newcommand{\algarray}{\opfont{array}}
\newcommand{\algcapacity}{\opfont{capacity}}
\newcommand{\algslotref}{\opfont{slot\_ref}}
\newcommand{\algcount}{\opfont{count}}
\newcommand{\algarraynew}{\opfont{array\_new}}
\newcommand{\algcapacitynew}{\opfont{capacity\_new}}
\newcommand{\algnull}{\opfont{null}}
\newcommand{\algcopied}{\opfont{copied}}
\newcommand{\algchangearray}{\opfont{change\_array}}
\newcommand{\algappend}{\opfont{append}}

\newcommand{\lastbind}{\it lastbind}

\newcommand{\ArrayDynAppendDereg}{\opfont{ArrayDynAppendDereg}}
\newcommand{\ArrayStatAppendDereg}{\opfont{ArrayStatAppendDereg}}
\newcommand{\ArrayStatSearchNo}{\opfont{ArrayStatSearchNo}}
\newcommand{\ArrayDynSearchResize}{\opfont{ArrayDynSearchResize}}

\newcommand{\fastcoll}{\opfont{FastCollect}}
\newcommand{\hohrc}{\opfont{HOHRC}}

\newcommand{\helpcopy}{\opfont{help\_copy}}
\newcommand{\attemptresize}{\opfont{attempt\_resize}}

\newcommand{\incoherent}{\color{black!30}}
\newcommand{\somewhatcoherent}{\color{black}}
\newcommand{\highlight}[1]{\textcolor{red}{[#1]}}

\newcommand{\vsp}{\vspace{.1in}}
\renewcommand{\paragraph}[1]{\noindent\textbf{\textrm{#1}}}

% Transaction types
\newcommand{\SendTx}{\opfont{SendTx}}
\newcommand{\CallTx}{\opfont{CallTx}}
\newcommand{\NameTx}{\opfont{CallTx}}
\newcommand{\BondTx}{\opfont{BondTx}}
\newcommand{\UnbondTx}{\opfont{UnbondTx}}
\newcommand{\DupeOutTx}{\opfont{DupeoutTx}}
\newcommand{\PermissionsTx}{\opfont{PermissionsTx}}

% Operations
\newcommand{\BroadcastTx}{\opfont{BroadcastTx}}

\newcommand{\proveMethod}{prove}
\newcommand{\claimMethod}{claim}
\newcommand{\verifySigMethod}{verifySig}
\newcommand{\respectsRulesMethod}{respectsRules}
\newcommand{\sendConfPolicyMethod}{sendConfPolicy}
\newcommand{\processMethod}{process}
\newcommand{\summaryMethod}{summary}
\newcommand{\ResetForBlockMethod}{ResetForBlock}

\newcommand{\proveMethodOp}{\opfont{\proveMethod}}
\newcommand{\claimMethodOp}{\opfont{\claimMethod}}
\newcommand{\verifySigMethodOp}{\opfont{\verifySigMethod}}
\newcommand{\respectsRulesMethodOp}{\opfont{\respectsRulesMethod}}
\newcommand{\sendConfPolicyMethodOp}{\opfont{\sendConfPolicyMethod}}
\newcommand{\processMethodOp}{\opfont{\processMethod}}
\newcommand{\summaryMethodOp}{\opfont{\summaryMethod}}
\newcommand{\ResetForBlockMethodOp}{\opfont{\ResetForBlockMethod}}

% Names
\newcommand{ \SenderName}{\ensuremath{S}}
\newcommand{ \ReceiverName}{\ensuremath{R}}

% Pseudocode defs

\newcommand{\ProtocolDelta}{\ensuremath{\mathit{GracePeriod}}}
\newcommand{\ProtocolHashFunction}{\ensuremath{\mathit{ripemd160}}}
\newcommand{\ProtocolConnID}{\ensuremath{\mathit{connID}}}

\newcommand{\SndrValues}{\ensuremath{\mathit{ValsSent}}}
\newcommand{\SndrLastSentIndex}{\ensuremath{\mathit{lastSent}}}
\newcommand{\SndrLastConfirmationIndex}{\ensuremath{\mathit{lastConf}}}
\newcommand{\SndrHashCache}{\ensuremath{\mathit{hashCache}}}
\newcommand{\SndrConfs}{\ensuremath{\mathit{confirmations}}}
\newcommand{\SndrReceiverChannel}{\ensuremath{\mathit{chan\ReceiverName}}}
\newcommand{\SndrHashResult}{\ensuremath{\mathit{\mathit{confHash}}}}

\newcommand{\RcvrMessagesReceived}{\ensuremath{\mathit{Msgs}}}
\newcommand{\RcvrRecordedEffects}{\ensuremath{\mathit{Data}}}
\newcommand{\RcvrLastReceivedIndex}{\ensuremath{\mathit{lastRcvd}}}
\newcommand{\RcvrLastAcknowledgedIndex}{\ensuremath{\mathit{lastAckd}}}
\newcommand{\RcvrPendingConfirmations}{\ensuremath{\mathit{pendingConfs}}}
\newcommand{\RcvrLastConfirmedIndex}{\ensuremath{\mathit{lastConf}}}
\newcommand{\RcvrCummulativeHash}{\ensuremath{\mathit{seqHash}}}
\newcommand{\RcvrSenderChannel}{\ensuremath{\mathit{chan\SenderName}}}

\newcommand{\TxMap}{\ensuremath{\mathit{TxMap}}}
\newcommand{\OutgoingQueue}{\ensuremath{\mathit{OutQueue}}}
\newcommand{\OutgoingOrder}{\ensuremath{\mathit{OutOrder}}}
\newcommand{\SummarySeq}{\ensuremath{\mathit{SummarySeq}}}
\newcommand{\Hash}{\ensuremath{\mathit{Hash}}}
\newcommand{\RecentBlock}{\ensuremath{\mathit{MostRecentCommittedBlock}}}

%
% compact labels and references

%
\newcommand{\figlabel}[1]{\label{figure:#1}}
\newcommand{\nakedfigref}[1]{\ref{figure:#1}}
\newcommand{\figref}[1]{Figure~\nakedfigref{#1}}
\newcommand{\figrangeref}[2]{Figures~\nakedfigref{#1}--\nakedfigref{#2}}

\newcommand{\linelabel}[1]{\label{line:#1}}
\newcommand{\nakedlineref}[1]{\ref{line:#1}}
\newcommand{\lineref}[1]{Line~\nakedlineref{#1}}
\newcommand{\linerangeref}[2]{Lines~\nakedlineref{#1}--\nakedlineref{#2}}

\newcommand{\seclabel}[1]{\label{sec:#1}}
\newcommand{\nakedsecref}[1]{\ref{sec:#1}}
\newcommand{\secref}[1]{Section~\nakedsecref{#1}}

\newcommand{\idtonly}[1]{}

% \conferenceinfo{TBD} {} 
% \CopyrightYear{2010}
% \crdata{}

\title{Enhancing Accountability and Trust in Distributed Ledgers\thanks{Copyright
    \textcopyright{} 2016, Oracle and/or its affiliates.  All rights
    reserved.}}

% This will be the format for the camera-ready version.
%
% \numberofauthors{1}
% \author{
% \alignauthor
% Bob, Fred, George and Reginald\\
%       \affaddr{Sun Microsystems Laboratories}\\
%       \email{\{bob,fred,george\}@sun.com}
% }

% This format works with sigplanconf.cls
%
%\authorinfo{First Author\superscript{\dag}\and SecondAuthor\superscript{\ddag}}
%           {{\dag}Institution One \and {\ddag}Institution Two}
%           {\footnotesize first.author@sun.com\and second.author@sun.com}

% This format works with the article documentclass
% 
\author{Maurice Herlihy \\
   Brown University and Oracle Labs \\
   mph@cs.brown.edu \\
   \and
   Mark Moir \\
   Oracle Labs \\
   mark.moir@oracle.com
}

\maketitle \thispagestyle{empty}

\begin{abstract}

  \emph{Permisionless} decentralized ledgers (``blockchains'') such as
  the one underlying the cryptocurrency
  Bitcoin % use proof of work (PoW) to
  allow anonymous participants to maintain the ledger, while
  avoiding control or ``censorship'' by any single entity.  In contrast,
  \emph{permissioned} decentralized ledgers exploit real-world trust and
  accountability, allowing only explicitly authorized parties to
  maintain the ledger. Permissioned ledgers support more flexible governance
  and a wider choice of consensus mechanisms.

  Both kinds of decentralized ledgers may be susceptible to manipulation
  by participants who favor some transactions over others.
  The real-world accountability underlying permissioned ledgers 
  provides an opportunity to impose fairness constraints that can be enforced by 
  penalizing violators after-the-fact.  To date, however, this opportunity has not
  been fully exploited, unnecessarily leaving participants latitude to manipulate
  outcomes undetectably.

  This paper draws attention to this issue, and proposes design principles
  to make such manipulation more difficult, as well as specific mechanisms to make it
  easier to detect when violations occur.

\end{abstract}

%\clearpage
%\setcounter{page}{1}
% \keywords{Accountability, trust, decentralized ledgers, blockchains}

\section{Introduction}
\seclabel{intro}
A \emph{blockchain} is a data structure used to implement tamper-resistant
distributed ledgers.
Multiple \emph{nodes} follow a
common protocol in which transactions from \emph{clients} are packaged into \emph{blocks},
and nodes use a consensus protocol to agree on successive blocks.  Each block's \emph{header} contains a
cryptographic hash of the previous block's header, making it difficult to tamper with the ledger.
Bitcoin~\cite{bitcoin} is the best-known blockchain-based
distributed ledger today.

% Distributed ledger implementations fall into two broad categories.
In \emph{permissionless} implementations, such as Bitcoin,
any node willing to follow the protocol can participate,
and anybody can generate addresses that can receive bitcoins, and 
can propose transactions that transfer bitcoins from any address for
which they have the associated private key.
By contrast, in \emph{permissioned} implementations,
% such as Stellar~\cite{stellar} and Factom~\cite{factom},
the sets of participating nodes are controlled by an authority,
perhaps one organization, perhaps a consortium.

A permissionless implementation makes sense for applications such as Bitcoin,
which seek to ensure that nobody can control who can participate,
a property often called \emph{censorship resistance}.
By contrast, permissioned implementations explicitly permit some forms of censorship:
for example, permitting compliance with ``know your customer'' regulations
that exclude known money-launderers from financial markets.
Moreover, permissioned implementations can often provide more effective governance:
for example, by providing an orderly procedure for updating the ledger
protocol~\cite{bitcoinfailed}.

Here, we focus on one more important difference:
permissioned ledgers can hold participants
\emph{accountable} for misbehavior in ways that permissionless implementations cannot.
Many distributed ledgers would benefit from \emph{fairness} guarantees.
For example, one client's proposed transactions should not be systematically
delayed longer than others',
or one client's transactions should not be systematically scheduled just after another's
competing transaction (say, front-running a stock purchase).
While it may be difficult or impossible to flag a single instance of a fairness
violation, systematic violations can be detected over time with high confidence.
We say that
a \emph{fairness policy} is a technique for ensuring a \emph{fairness guarantee}.

In permissionless ledgers, such as Bitcoin,
an inherent lack of accountability makes fairness policies difficult to enforce.
For example, in a \emph{mining cartel attack}~\cite{CourtoisB14,eyal2014selfishMining,cartel},
a group of Bitcoin miners ignores transactions proposed by miners which are
not members of that group.
It can be difficult to detect such behavior,
and even if detected, the cartel members suffer no effective penalty,
other than perhaps public shaming and a loss of confidence in the ledger itself.

Even in permissioned ledgers, however,
a participating node may violate a fairness policy because it has been hacked,
its software is defective, its operator is dishonest, and so on.
\emph{In principle},
permissioned ledgers make it easier to hold nodes accountable for
fairness policy violations:
once exposed, a violator may lose a deposit,
it may be expelled from the ledger, or it may be sued.
In practice, however, reducing the opportunities for internal fairness
violations, and detecting them when they occur,
is a non-trivial problem that requires systematic scrutiny.

Our contribution is to call attention to this issue,
and to propose principles for designing permissioned decentralized ledgers to
hold participants accountable for fairness policy violations.
To demonstrate that these principles apply to real systems,
we propose a number of modifications to the open-source
Tendermint~\cite{tendermint-github} ledger system.
Tendermint provides an
effective platform for illustrating our ideas, but we believe our principles are 
applicable to a wide range of ledger implementations.
We hope that both the general principles 
and the specific techniques we introduce
will serve to alert the community to the importance of ensuring fairness
properties in distributed ledgers,
and to contribute to effective ways to do so.

Our approach is based on the following design principles.
\begin{itemize}
\item
Each non-deterministic choice presents an opportunity to manipulate outcomes.
As much as possible, non-deterministic choices should be replaced
with deterministic rules,
and mechanisms for detecting violations and holding the culprits accountable
should be available.

\item
Non-deterministic choices that cannot be avoided should be disclosed in an
auditable way, 
ensuring that they cannot be altered later,
and that unusual statistical patterns can be detected.

\item
As much as possible,
mechanisms that enhance accountability should be kept off the system's
critical path in order to avoid imposing substantial costs on
normal-case execution by honest participants.
\end{itemize}

Sometimes, participants can be held accountable right away:
posting a proof that a participant cheated may cause that participant to be expelled.
%(most distributed ledgers have such a mechanism for egregious violations). 
Often, participants can be held accountable only after-the-fact.
For example,
sometimes one can detect that one of two interacting parties cheated,
but it may not be immediately clear which one.
Sometimes individual violations may be indistinguishable from legitimate non-determinism.
In both cases, reducing opportunities for non-deterministic choices
and systematically logging remaining ones can reveal suspicious long-term trends.

Section~\ref{sec:background} presents background on the
most relevant parts of Tendermint~\cite{tendermint-github}, the open-source
distributed ledger system on which we base our design.
\secref{design-overview} overviews our design,
details of which are presented in
\secref{detailed-description}.  Discussion and related work appear 
in Sections \nakedsecref{discussion} and \nakedsecref{related}, and we conclude in
\secref{concluding-remarks}.

\section{Background on Tendermint}
\seclabel{background}

In this section, we survey aspects of Tendermint~\cite{tendermint-github} used
to embody our ideas.\negthinspace\footnote{Our extensions build on commit
  \texttt{c318a227}   from October 3, 2015. Since
  then, work on Tendermint has progressed, but as far as we can tell,
  later versions do not substantially change the
  issues addressed in this paper.  Nonetheless details of integrating techniques
  we describe may differ given recent architectural 
  changes in Tendermint \cite{Tendermint-TMSP}.}%
More complete descriptions are available elsewhere \cite{tendermint-wiki}.

Tendermint employs a peer-to-peer network in which nodes gossip
information including: \emph{transactions} submitted by clients,
\emph{blocks} of transactions proposed by \emph{proposers}\negthinspace,
and various messages used by the consensus protocol used to  reach
agreement among nodes on successive \emph{next} blocks in the \emph{chain}
(blockchain).
Honest validators vote only for blocks that satisfy certain validity conditions:
for example, each block must include
a cryptographic hash of the previous block,
making it essentially impossible to change a block already committed into the blockchain.

% \item[Peer-to-peer structure}]
% Upon starting, a peer can specify the address of a ``seed'' node,
% allowing it to join the network and begin receiving committed
% previously committed blocks in a ``fast sync'' mode. When it ``catches
% up'' to the rest of the network, it begins participating in the
% protocol proper. Nodes periodically request information about other
% nodes from their immediate peers, allowing them to add new peers as
% needed.

Tendermint
supports a notion of \emph{accounts}, each of which has some
number of associated \emph{tokens}.
Unlike Bitcoin, tokens are not created by proof-of-work mining.
Instead, they are created by the first, or \emph{genesis} block,
or acquired out-of-band, perhaps by purchase.

Tendermint's consensus mechanism is a variant of the PBFT Byzantine
agreement algorithm \cite{PBFT,tendermint-vs-pbft}.
The current blockchain length is called its \emph{height}.
To add a new block of transactions to the chain,
the protocol executes a sequence of \emph{rounds}.
At each height and round, a \emph{proposer} proposes a block,
and \emph{validator} nodes vote whether to accept it.
Further details are unimportant for our purposes:
we use Tendermint's consensus mechanism unmodified,
and the techniques we present are not specific to the consensus protocol.

Tendermint uses a \emph{proof of stake} (PoS) mechanism to keep
validators honest.
% The proposer for a given height and round is chosen deterministically.
Each validator posts a bond, in the form of tokens,
which it will lose if it is caught violating the protocol.
If a validator is caught,
proof of that transgression is posted to the blockchain,
the culprit is expelled and its bond confiscated.
(Others have discussed the strengths and weaknesses of PoS consensus~\cite{nothing-at-stake}.)

Here we are less concerned with the specific mechanisms used to punish
participants that violate fairness constraints
than how to reduce opportunities for such violations,
and how to detect them when they occur.

Tendermint uses peer-to-peer gossiping to propagate consensus-related
information including block proposals, votes, and other state information.
Tendermint propagates block data in \emph{parts},
using a protocol inspired by LibSwift~\cite{libswift},
in which each block's header contains the root of a Merkle tree~\cite{Merkle:1987:DSB:646752.704751}.
The Merkle tree contains hashes of each block part,
allowing confirmation that received block parts are valid.

The mechanisms described in this paper are independent of these
aspects of Tendermint, with minor exceptions mentioned later.
Next, we discuss in more detail several aspects of Tendermint that are
more directly relevant to the mechanisms we present.

We focus here on the two transaction types of most direct interest to
users. \SendTx{} transactions send tokens from one or more \emph{input}
accounts to one or more \emph{output} accounts.
\CallTx{} transactions create \emph{contracts},
expressed as Ethereum \cite{ethereum} virtual machine byte code.
Each such transaction establishes an address for the contract, enabling
subsequent \CallTx{} transactions to invoke methods of the contract.
Contracts are special accounts that have an associated key-value store,
which can be modified by the contract's methods.
Like \SendTx{} transactions, \CallTx{} transactions have at least one
input, which can be used to send tokens to be stored
in the contract's account.
Contract methods can in turn send tokens from the contract's balance to other
accounts (either contracts or regular accounts).

To prevent ``replay attacks''\negthinspace,
in which an attacker could cause a payment to be made twice,
each transaction input includes a sequence number known as a \emph{nonce}.
An input is valid only if its nonce is one greater than that of
the input from that account in the previous transaction that spends from that account.

A transaction enters the network via a \BroadcastTx{} RPC
call to a node; we call it an \emph{acceptor} node.
Each node has a local \emph{mempool} that stores transactions the
node has received but has not yet seen included in a committed block.
A node validates each transaction before adding it to its mempool.
Validation verifies that each of the transaction's inputs have valid signatures,
that its account has sufficient balance,
and that the nonce is correct.
The balance and nonce validations are performed against the state achieved by
applying the acceptor's mempool transactions to the last committed block the acceptor has seen.

A node receiving a transaction from a peer validates it and appends it to its mempool,
as described above.
A node has a ``peer broadcast'' routine for each of its peers, which periodically 
(every 1ms by default) takes a snapshot of the mempool, and gossips transactions from
this snapshot to the peer in the order they appear; preserving this
order ensures that nonce values remain in correct sequence.

When a node receives a newly committed block, it removes
from its mempool all transactions included in the block
and transactions invalidated by the new state (for example,
a transaction is removed if its nonce is no longer correct); it then communicates these changes
to each peer broadcast routine to suppress unnecessary late gossiping.

The proposer for a given height and round proposes a prefix of its mempool
transactions, in mempool order, thereby preserving correct nonce order 
of transactions spending from the same account.

\subsection{Opportunities to Violate Fairness}

\paragraph{Censorship}
A proposer could pretend it never received a transaction,
either directly from the client, or indirectly via gossip\vsp.

\paragraph{Order manipulation}
A correct Tendermint proposer proposes transactions within a block in the
local mempool order.
Nevertheless,
a dishonest proposer could reorder transactions without fear of detection,
because that proposer's mempool order is not externally visible.
Indeed, the Tendermint wiki~\cite{tendermint-vs-pbft} observes that
`the power to re-order transactions is shared equally among the participants''\negthinspace\vsp.

\paragraph{Transaction Injection}
After receiving a transaction from a peer,
a proposer (or its ally) could create another transaction and order it first.
For example,
a node that has been bribed by an unscrupulous hedge fund might detect that
a pension fund has placed a stock order.
That node could inject a buy order in front of the pension fund's,
immediately reselling that stock to the pension fund at a markup,
a practice known as \emph{front-running}.

%%% Local Variables: 
%%% mode: latex
%%% TeX-master: t
%%% End: 

\section{Design Overview}
\seclabel{design-overview}

This section gives an overview of our extensions to the Tendermint protocol; 
Section~\ref{sec:detailed-description} presents details.

Ideally, a proposer should not be able to (1) pretend it has not
received a transaction that it received,
(2) control which received transactions to include in its proposed block,
(3) control their order in the block,
or (4) inject transactions ahead of those it has received.
Any such attempt should preferably result in undeniable proof of misbehavior,
and at least produce evidence that can be accumulated and analyzed after-the-fact.

We can make fairness violations
much more difficult to achieve without detection by imposing
deterministic rules governing the order in which transactions are
propagated (making ``missing'' transactions apparent), which 
transactions are included in each proposed
block, and the order in which they appear.
These forms of accountability are achieved by
requiring nodes to regularly disclose auditable state information\vsp.

\paragraph{Transaction Ingress}
When a Tendermint node accepts a transaction via a \BroadcastTx{} RPC
call, it returns a \emph{receipt} to the caller.
We augment this receipt to include \emph{acceptor data},
which includes an \emph{acceptor ID},
a \emph{sequence number}, and an \emph{acceptor signature} that covers
both the transaction and the acceptor ID and sequence number.
This data gives the caller undeniable proof that the transaction has been
accepted into the network; together with additional protocol
extensions described below, it also enables detection of certain attempts
to censor the transaction or manipulate its ordering.
We now concentrate on handling transactions that have been accepted\vsp.

\paragraph{Accountable Legitimate Censorship}
When a node has a legitimate reason for censoring a transaction
(for example, it spends from an account on a government blacklist),
the node cannot simply drop the transaction.
Instead, it explicitly invalidates the transaction by appending signed metadata
indicating that this transaction is being censored,
perhaps including a justification.
The explicitly invalidated transaction is forwarded as usual,
and is eventually put in a block and committed the the chain,
making the censorship procedure transparent and accountable\vsp.

\paragraph{Transaction Propagation}
Recall that an acceptor validates each transaction against its most recent
committed block followed by the earlier uncommitted transactions it has accepted.
In contrast to Tendermint, the transaction is \emph{not} validated again during the gossip protocol.
Instead, it is validated again only when a proposer includes it in a block (either processed or explicitly rejected).
The initial validation ensures that the transaction can execute successfully in
at least \emph{some} scenario.
The absence of intermediate validation ensures that a node propagating a
transaction has no excuse to drop that transaction,
ensuring that every such transaction will eventually reach a proposer.
The proposer's final validation ensures that the sequence of transactions 
included in its proposed block can 
execute successfully starting from the state after the previous block\vsp.

Tendermint's mempool serves several purposes: (1) to
represent a state against which incoming transactions are validated, (2) to
order transactions for gossiping, and (3) to select transactions
for proposals.
We split the mempool into two distinct pieces to reflect this functionality.
An acceptor's \emph{accepted transaction queue} holds the sequence of
transactions it has accepted but have not yet been included in a block.
Newly-submitted transactions are validated against the state achieved by starting with
the
previously committed block and executing each of the transaction's predecessors in that queue in order.
The \emph{outgoing queue} orders transactions for gossip and for inclusion in proposals
(Section~\ref{sec:usingoneway}).  This separation supports the above-described modification by
allowing an acceptor to validate transactions it accepts only against previous transactions it
has accepted, rather than against all gossiped transactions as in Tendermint.

To minimize opportunities for nodes to drop or reorder transactions,
we impose the following rule:

\begin{description}
\item[Transaction propagation rule]
When a node accepts a transaction from a client or receives one from a peer,
it must send that transaction to each of its peers, 
unless the transaction has already been included in a committed block.
All such transactions must be sent in the same order to each peer,
in the same order they were received from each peer,
and preserving the acceptor ID order for transactions from the same acceptor.
There is one exception:
a transaction received from multiple peers should be included in the outgoing order once only,
and later duplicates are dropped.
\end{description}

The next section proposes mechanisms for making nodes accountable for following this rule,
as well as for transaction selection and ordering in proposals.

\section{Detailed Description}
\seclabel{detailed-description}
This section gives a more detailed account of the changes to Tendermint
proposed to enhance accountability for fairness violations.
Tendermint already follows the Transaction Propagation Rule presented
in the previous section, but there is no mechanism for detecting
violations. Furthermore, nodes can easily exclude or reorder
transactions without detection. \secref{gossip-accountability}
requires nodes to regularly disclose concise summaries of their
outgoing queues, which can subsequently be audited. Moreover, to deny
proposers power to select and order transactions, each proposal must
include proof that deterministic rules were followed to select
transactions from the proposer's outgoing queue
(\secref{choosing}) and to order them in the proposal
(\secref{ordering}).

\subsection{Making Gossip Order Accountable}
\seclabel{gossip-accountability}

Detecting violations of the Transaction Propagation Rule requires reliably
recording nodes' internal states and communication states.
To address this issue, we propose a novel \emph{one-way accountable channel}
mechanism which provides a building block for accountable one-way communication
from one node to another.

\subsubsection{One-way Accountable Channel (OWAC)}

Suppose a sender \SenderName{} sends a sequence of values to a
receiver \ReceiverName{}, which is required to process them in the order sent.
Ideally, when \SenderName{} sends a value to \ReceiverName{},
it would then be able to prove that it had done so,
requiring \ReceiverName{} to process values in the order sent,
with no opportunity to examine a message and then pretend not to have received it.
Using a classical cryptographic \emph{commitment}
protocol~\cite{Brassard:1988:MDP:53813.53817} 
would require \ReceiverName{} to acknowledge a message's hash
before receiving the message itself, but does not solve the problem:
\ReceiverName{} can still pretend not to have received a message,
even after acknowledging its hash.
It would also require an additional round-trip for each message,
which is unacceptable in a gossip protocol.

\begin{figure*}[h]
\begin{minipage}[t]{3.25in}
\begin{lstlisting}[language=pseudo,escapeinside={@}{@},mathescape,name=oneway,basicstyle=\tiny]
Constants

@\ProtocolDelta@: integer  // max. confirmation @``@lag@''@
@\ProtocolConnID@: connection ID   // to avoid replay attacks



Sender @\SenderName@

@\SndrValues@: array of values
@\SndrLastSentIndex@: integer = -1    // last sent index
@\SndrLastConfirmationIndex@: integer = -1   // last confirmed index
@\SndrHashCache@: integer -> hash = {(-1,0)}
@\SndrConfs@: integer -> signed message = {}
@\SndrReceiverChannel@: secure FIFO channel to @\ReceiverName{}@

boolean send$(v)$
@\linelabel{sndr:enoughconfs}@    if $\SndrLastSentIndex > \SndrLastConfirmationIndex +  \ProtocolDelta$
@\linelabel{sndr:denied}@      return false // confirmations too far behind
    $\SndrLastSentIndex++$
    $\SndrValues[\SndrLastSentIndex] = v$
@\linelabel{sndr:msg}@    $m = \langle \ProtocolConnID, \SndrLastSentIndex, v, \SndrLastConfirmationIndex\rangle$
@\linelabel{sndr:send}@    @\SndrReceiverChannel@.send$(\mathit{sign}_\SenderName(m))$
    return true

when Receive $m =\langle id, k, \SndrHashResult ,s\rangle$
    if not verifySig($m$)
        @\claimMethod@(@``@invalid signature@''@)
    if $id \neq$@\ProtocolConnID@
        @\claimMethod@(@``@invalid connection ID@''@)
    if  $k \leq \SndrLastConfirmationIndex$ 
        @\claimMethod@(@``@confirmation out of sequence@''@)
    if  $k > \SndrLastSentIndex$ 
        @\claimMethod@(@``@cannot confirm unsent values@''@)
@\linelabel{sndr:hashbegin}@    $j, h = \SndrLastConfirmationIndex, \SndrHashCache[\SndrLastConfirmationIndex]$
@\linelabel{sndr:hashloopbegin}@    while $j < k$
@\linelabel{sndr:hashloopend}@        $j++$
@\linelabel{sndr:hash}@        $h = \ProtocolHashFunction(h \cdot{ } \SndrValues[j])$    
@\linelabel{sndr:hashcache}@    $\SndrHashCache[k] = h$
    if $\SndrHashResult \neq h$
        @\claimMethod@(@``@incorrect hash confirmation@''@)
    $\SndrLastConfirmationIndex = k$
@\linelabel{sndr:recordconf}@    $\SndrConfs[\SndrLastConfirmationIndex] = m$
\end{lstlisting}
\end{minipage}
\
\begin{minipage}[t]{3.25in}
\begin{lstlisting}[language=pseudo,escapeinside={@}{@},mathescape,name=oneway,basicstyle=\tiny]
Receiver @\ReceiverName@

@\RcvrMessagesReceived@: integer -> signed message = {}
@\RcvrRecordedEffects@: recorded effects of processed messages
@\RcvrLastReceivedIndex@: integer = -1  // received index
@$\RcvrLastConfirmedIndex$@: integer = -1  // last confirmed index
@\RcvrLastAcknowledgedIndex@: integer       // last acknowledged index
@\RcvrCummulativeHash@: hash = 0      // hash of value sequence
@\RcvrSenderChannel@: secure FIFO channel to @\SenderName{}@

when Receive $m = \langle id, i,v,k\rangle$
    if not @\verifySigMethod@($m$)
        @\claimMethod@(@``@invalid signature@''@)
    if $id \neq$@\ProtocolConnID@
        @\claimMethod@(@``@invalid connection ID@''@)
@\linelabel{rcvr:ackseqchk}@    if $k < \RcvrLastAcknowledgedIndex$
        @\claimMethod@(@``@acknowledgement out of sequence@''@)
    if $k > \RcvrLastAcknowledgedIndex$
        if $k \neq \RcvrPendingConfirmations.removeMin()$
            claim(@``@invalid acknowledgement@''@)
@\linelabel{rcvr:setlastack}@        $\RcvrLastAcknowledgedIndex = k$      
@\linelabel{rcvr:checkseqhigh}@    if $i > \RcvrLastReceivedIndex + 1$ 
@\linelabel{rcvr:claimseqskip}@        @\claimMethod@(@``@skipped message@''@)
@\linelabel{rcvr:checkseqlow}@    if $i < \RcvrLastReceivedIndex + 1$ 
@\linelabel{rcvr:checkdup}@      if $\RcvrMessagesReceived[i] = m$
@\linelabel{rcvr:claimdup}@        @\claimMethod@(@``@duplicate message@''@)
      else
@\linelabel{rcvr:provecheating}@        @\proveMethod@(@``@conflicting messages@''@$, m, \RcvrMessagesReceived[i]$)
    $\RcvrLastReceivedIndex = i$
@\linelabel{rcvr:recordmsg}@    $\RcvrMessagesReceived[\RcvrLastReceivedIndex] = m$
    if $\RcvrLastReceivedIndex > \RcvrLastAcknowledgedIndex + \ProtocolDelta$
        @\proveMethod@(@``@too far ahead@''@, $\RcvrMessagesReceived, \RcvrLastReceivedIndex$)
    if not @\respectsRulesMethod@($\RcvrMessagesReceived, \RcvrLastReceivedIndex$)
        @\proveMethod@(@``@rules violated@''@, $\RcvrMessagesReceived, \RcvrLastReceivedIndex$)
@\linelabel{rcvr:hash}@    $\RcvrCummulativeHash = \mathit{ripemd160}(\RcvrCummulativeHash \cdot{ } v)$
@\linelabel{rcvr:process}@    @\processMethod@($\RcvrRecordedEffects, v$)  
    
@\linelabel{rcvr:event}@when Event     // timer or other event
@\linelabel{rcvr:confpolicy}@    if @\sendConfPolicyMethod@($\RcvrLastAcknowledgedIndex, \RcvrLastConfirmedIndex, \RcvrLastReceivedIndex$)
@\linelabel{rcvr:message}@        $m = \langle \ProtocolConnID, \RcvrLastReceivedIndex, \RcvrCummulativeHash, \summaryMethod(\RcvrRecordedEffects) \rangle$
@\linelabel{rcvr:send}@        C.send$(\mathit{sign}_\ReceiverName(m))$
        $\RcvrLastConfirmedIndex = \RcvrLastReceivedIndex$
@\linelabel{rcvr:recordconf}@        $\RcvrPendingConfirmations.insert(\RcvrLastReceivedIndex)$
\end{lstlisting}
\end{minipage}

\caption{Protocol in which Sender sends sequence of values to Receiver, and Receiver confirms messages correctly recieved.  Confirmations may lag sends by up to \ProtocolDelta{} messages, but sending does not continue after that unless and until confirmations catch up within \ProtocolDelta.  Protocol violations detected are reported via \claimMethodOp{} when undeniable proof can't be provided, and \proveMethodOp{} when it can; in the latter case, data on which proof may be based are included as arguments.  Each method executes atomically.}
\figlabel{one-way}
\end{figure*}

Our goal is a pragmatic design that does not overly burden common-case
execution,
but that yields proof of deviation when possible,
and otherwise yields evidence of possible deviation that can be aggregated over
time with other evidence.
Detailed pseudocode illustrating the OWAC protocol appears in 
\figref{one-way}.

The protocol description serves two purposes.
First, it \emph{specifies} how correct OWAC implementations can behave.
Second, it illustrates how \SenderName{} and \ReceiverName{} can detect and
react to protocol violations by the other.
If one party can prove that the other has violated the protocol,
it calls \proveMethodOp{}, with the proof as argument.
If one party can detect (but not prove) that the other has violated the protocol,
it instead calls \claimMethodOp, thereby contributing evidence for potential investigation.
Each party logs the messages it receives (\lineref{sndr:recordconf} and \lineref{rcvr:recordmsg}),
which can be used to support, defend against, or help evaluate allegations of misbehavior.

Validity rules are modeled by a \respectsRulesMethodOp{} method.
Similarly, \sendConfPolicyMethodOp{} represents a policy that decides when
\ReceiverName{} should confirm messages received since its last confirmation,
\processMethodOp{} encapsulates how \ReceiverName{} processes messages,
and \summaryMethodOp{} produces a concise summary of the result.

For brevity, we elide details such as when
historical data can be deleted
% how a new OWAC is established starting from an index other than 0 in the
% sequence of values,
and what happens after one party accuses the other of violating the protocol.

The key idea behind the OWAC protocol is that each party computes,
for each message sent,
a cryptographic hash based on that message and prior messages.
These hashes are cumulative (the hash for each message is obtained by
concatenating the message to the hash of the previous message, and
hashing the result; see Lines \nakedlineref{sndr:hash} and \nakedlineref{rcvr:hash}).
In this way, a single hash summarizes the sequence of messages received up to some point.
Specifically, \ReceiverName{} sends confirmation of the messages it has
received so far, including the hash of the last message received
(\linerangeref{rcvr:message}{rcvr:send}). \SenderName{}
checks that the confirmation contains the correct hash for the
messages confirmed
(\linerangeref{sndr:hashbegin}{sndr:hashcache}), thus
obtaining proof that \ReceiverName{} has   received exactly
the messages sent up until that point.
\SenderName{} records these confirmations in case \ReceiverName{}
subsequently denies having received the confirmed messages
(\lineref{sndr:recordconf}).

At any time (\linerangeref{rcvr:event}{rcvr:confpolicy}),
\ReceiverName{} can confirm receipt of a \emph{batch} of messages
(\linerangeref{rcvr:message}{rcvr:send}),
allowing confirmations to be piggybacked onto messages already sent in
the host application (here, Tendermint).  
Each time \SenderName{} receives a confirmation,
it uses the hash calculated in response to the previous batch of
  confirmations (\lineref{sndr:hashbegin}) to calculate and store the hash for this batch
(\linerangeref{sndr:hashloopbegin}{sndr:hash});
this facilitates   verification of the next confirmation 
from \ReceiverName{} (\lineref{sndr:hashcache}).
Correspondingly, \ReceiverName{} records the highest   index of each batch of
messages it confirms (\lineref{rcvr:recordconf}), so
it can check that \SenderName{} correctly acknowledges its
 confirmations in order (\linerangeref{rcvr:ackseqchk}{rcvr:setlastack}).

This batching mechanism
allows confirmations from \ReceiverName{} to lag behind messages
sent by \SenderName{}.
To ensure that \ReceiverName{} cannot ignore a message without eventually
being detected, \SenderName{} allows at most \ProtocolDelta{} messages to be
outstanding without confirmation before it refuses to send additional messages
(\linerangeref{sndr:enoughconfs}{sndr:denied}). 
If \ReceiverName{} ignores a message,
it essentially shuts down that channel, thereby ensuring detection.  

In addition to confirming messages received, \ReceiverName{} 
processes each message (\lineref{rcvr:process}) and includes with its
confirmation a summary of its local data (\lineref{rcvr:message}), 
allowing it to prove it has indeed processed all messages received if challenged.
Similarly,
\SenderName{} includes with each message the  highest index for which it has
received confirmation (\lineref{sndr:msg}),
so that \SenderName{} cannot subsequently accuse \ReceiverName{} of
failing to confirm messages\vsp.

\paragraph{Accusations and Proofs}
In most cases, when \SenderName{} or \ReceiverName{} detects a violation,
the proof is self-explanatory.
In some cases,
one party can detect that the other party violated the protocol,
but cannot prove it to a third party.
In such cases,
that party calls \claimMethodOp{} rather than \proveMethodOp{}.
For example, if \SenderName{} were to skip a message (perhaps with the
intent of accusing \ReceiverName{} of ignoring a message it sent),
\ReceiverName{} would detect the omission (\lineref{rcvr:checkseqhigh}),
but would be unable to prove that \SenderName{} did not send
the skipped message.
Nevertheless, \emph{asserting} that the other party has misbehaved adds
to a body of evidence that may establish a pattern of
misbehavior.

In contrast, in some cases, a protocol violation can be proved.
For example, if messages sent by
\SenderName{} violate application-specific rules encapsulated in \respectsRulesMethodOp{},
\ReceiverName{} can produce a proof (because the messages are signed by
\SenderName{} and cannot be forged).
If \SenderName{} resends a previous message,
\ReceiverName{} cannot prove it,
and therefore   simply asserts a protocol violation (\lineref{rcvr:claimdup}).
On the other hand, if \SenderName{} sends \emph{different} messages
for the same index, \ReceiverName{} can prove it did so (\lineref{rcvr:provecheating}).

We emphasize that the primary purpose of enabling \SenderName{} and
\ReceiverName{} to monitor one another's compliance is to discourage
noncompliance; therefore violation reports should be rare,
especially provable ones.

\subsubsection{Using the One-Way Channel in Tendermint}
\seclabel{usingoneway}
To use the OWAC for gossiping transactions in Tendermint,
we instantiate \respectsRulesMethodOp{} with a method that performs
basic validation of transactions and also ensures that each node propagates
transactions from each acceptor in acceptor order, thus preventing nodes
from dropping or reordering transactions. Nodes are motivated to
verify that transactions received from each of its peers are valid and
preserve acceptor order before passing transactions on to their peers;
otherwise, they may be blamed for others' transgressions.

Confirmations are piggybacked on consensus-related messages in
Tendermint, in which a node informs its peers each time it enters a
new consensus round or changes steps within a round (that is, the
``event'' that triggers confirmations is when the node is sending a
\texttt{RoundStep} message anyway).

The \processMethodOp{} method executed upon receiving a transaction from a peer ensures that
the transaction is in the node's outgoing queue.  (When a node accepts a transaction
from a client, it inserts the transaction into the outgoing queue in the same way after adding
acceptor data.) The \summaryMethodOp{} method
returns a concise summary of the node's outgoing queue,
along with other information described below,
allowing the node to subsequently prove that it indeed has the transaction in
its outgoing queue (until the transaction is included in a block).
The summary serves two purposes: first, it supports gossip accountability by
allowing the node if challenged to prove that it is faithfully
following the protocol when gossiping transactions, and second, it
allows a proposer to prove that the set of transactions included in
its proposed block is consistent with the transaction selection rules
discussed in \secref{choosing}\vsp.

\paragraph{Gossip Accountability}
The summary of the
outgoing queue 
provided by the \summaryMethodOp{} method 
includes transactions received from \emph{all} peers
(thus, the \processMethodOp{} and \summaryMethodOp{} methods for all
incoming OWACs on a node share a common
outgoing queue). Furthermore, summaries include the height and hash of
the most recently seen committed block, and a sequence number that
orders summaries produced by the same node. The implied
ordering allows the outgoing queues represented by two summaries to be
compared (even if the two summaries were sent to different peers) to
verify that the node did not ``lose'' or reorder transactions in its
outgoing queue, and ensuring that consistent information is being sent
to all peers.

To support these accountability mechanisms, a node's outgoing queue is 
represented as a key-value store implemented using a
dynamic Merkle tree \cite{Merkle:1987:DSB:646752.704751}.
The keys are integers that order
transactions in the outgoing queue, and the values are transactions.
(Tendermint includes two Merkle tree implementations. The one used for
verifying that the parts of a block's data have been received
correctly, as discussed earlier, is for static trees. For the outgoing
queue, we use the more sophisticated Merklized AVL+ tree, which
supports dynamic inserts and deletes and is used in Tendermint for
purposes such as recording accounts and storage.)

Representing the outgoing queue as a Merkle tree allows each node to provide
concise summaries of ``snapshots'' of its outgoing queue at various
times, and to prove the accuracy of subsequent responses to requests
for more detailed information about the content of the outgoing queue
at those times. \secref{accountability} discusses how this enables accountability mechanisms.

The Merkle tree representation of the outgoing queue supports
accountability, but does not efficiently
support other outgoing queue operations.
For this reason, a node also stores its outgoing queue transactions in
additional structures, for internal use only,
that do not require the same level of accountability.

Specifically, transactions in the outgoing queue are also stored in
order in an array, exactly as in Tendermint's mempool (see
\secref{background}), except that a transaction that becomes
invalid due to the inclusion of another transaction in a
newly-committed block is not removed, because it should continue
propagating until a block is committed that explicitly rejects the transaction.  As in
Tendermint, this representation allows peer broadcast routines to
efficiently snapshot the outgoing queue, as well as allowing a node to
efficiently enumerate transactions to be included in a proposal block.

A local hashmap is also used to map transactions to ordering keys,
thereby facilitating duplicate transaction quashing and removal 
of transactions
included in a newly-committed block
from the Merkle tree.

\idtonly{Persons skilled in the art, given benefit of our description
  and the open-source Tendermint code, will readily implement these
  ideas as described.  They will also appreciate a range of
  alternative ways to achieve the same goals: i) complying with the
  Transaction Propagation Rule (\secref{design-overview}), ii)
  providing concise summaries of its outgoing queue, and iii) using
  these summaries to prove that details provided later are consistent
  with the summaries.}

\subsubsection{Accountability}
\seclabel{accountability}
Who performs accountability checks, how often, and why,
are beyond the scope of this paper.
Instead, we focus on how the mechanisms described
help detect violations of the Transaction Propagation Rule.

As described already,
whenever a node confirms a batch of received transactions,
it includes the root hash of the Merkle tree representing its outgoing queue.
If the node is challenged to prove that a given transaction was
included in the outgoing queue as of a certain summary,
or even to provide the queue's entire contents,
it can prove the accuracy of its response using standard Merkle proof
techniques~\cite{Merkle:1987:DSB:646752.704751}.

A peer can compare the transactions it has sent to a node
against such snapshots of the node's outgoing queue to detect
if the node drops or reorders transactions. The block
height and hash included in summaries enable confirmation
that transactions are dropped only when they are included in a block.

If a node is found to have cheated,
say by dropping or reordering transactions in the outgoing queue,
the summaries and contents can provide proof that the node has violated the rules.
If a node is unwilling or unable to respond to a challenge,
it can be penalized using built-in mechanisms such as 
forfeiting tokens or being excluded, as well as external mechanisms such as lawsuits, regulatory
penalties, and reputational harm.
How long a node is required to retain data to facilitate responses to challenges is a policy question.

\subsection{Selecting Transactions for a Proposal}
\seclabel{choosing}

The mechanisms that support outgoing queue accountability can also
limit proposers' power to select transactions.
The Transaction Propagation Rule (\secref{design-overview}) implies that
a node can choose any prefix of its outgoing queue to form a block proposal
that satisfies the following constraint: for any transaction in the proposal, 
all previous transactions
from the same acceptor have either already been included in a
committed block, or are also included in the proposal.

The length of the proposed prefix is a policy issue.
A deterministic rule substantially reduces opportunities for
manipulation by proposers.
One could choose a constant length, or the longest prefix not exceeding a fixed
transaction size total.
Relevant parameters could be fixed at creation time (in the ledger's genesis block),
adjusted deterministically (like Bitcoin's difficulty level),
or adjusted by the ledger's governing body.  Deterministic
selection makes manipulation such as front running
significantly more cumbersome,
easier to detect, and easier to assign blame, but does not make it impossible.
Stronger, but slightly
more intrusive, measures are possible; see \secref{discussion}.

To respond to possible future challenges about whether a proposal complies with the
deterministic rule,
each proposer includes in each proposal its outgoing queue's Merkle root.
Even greater accountability can be achieved by requiring verification 
\emph{before} committing the block.

Tendermint's mechanism for communicating blocks and transactions
(see \secref{background})
can be extended to include a representation of the outgoing queue.
Some simple optimizations can largely avoid the increase
in bandwidth and storage costs resulting from a naive implementation.
First, note that the block need not include the
entire contents of the outgoing queue. It suffices to include
enough of the outgoing queue to cover all selected transactions, plus
additional hashes to enable verification that these
transactions indeed form the correct prefix of the outgoing queue.

For example, if only the left subtree leaves were included,
we could include a representation of the left subtree that would prove that
the selected transactions form an appropriately-sized prefix of the outgoing queue,
together with the root hash of the right subtree.
This information is enough to verify that the selected transactions were a prefix of the
outgoing queue.
Here is how to generalize this example.
Tendermint's Merkle AVL+ tree has an in-order traversal method that
applies a provided function to each node visited. The
function returns an indication of whether the
traversal should stop. A simple modification of this mechanism allows
traversal to continue after enough transactions have been collected,
but to refrain from recursing into each new subtree encountered,
instead yielding the hash of the subtree's root. 

The result is a disclosure of the outgoing queue that
suffices to verify that the selected transactions are a prefix of the
outgoing queue, and that the Merkle root of the outgoing queue matches
the one included in the block header.
Tree size is linear in the number of \emph{selected} transactions,
and is independent of
the total number of transactions in the outgoing queue.

Since only proposal transactions (and not the
entire outgoing queue) are included,
the partial Merkle tree can replace transactions with their indexes in the block.
Validators need not persist the partial Merkle tree after verification.

\subsection{Ordering Proposal Transactions}
\seclabel{ordering}

Nodes can control the order in which transactions received
from different peers at around the same time are added to their outgoing
queues.
To avoid abuse of this ability,
a proposer is required to reorder transactions in a way that nobody can influence and everyone can verify.
Simply randomizing the order could violate dependencies between valid transactions
that spend from the same account,
resulting in rejection due to incorrect nonce values.

We emphasize that while proposers must preserve the order of transactions
that spend from the same account,
there is \emph{no} requirement to preserve acceptor order.
Acceptor order serves only to ensure that transactions cannot surreptitiously
be dropped in transit.

There are many ways to ensure that the order in which transactions are proposed
for commitment into the blockchain cannot be unfairly manipulated.
Here is one.
First, order the transactions in a random order that is beyond anyone's control and
that everyone can verify.
A simple and efficient way is to use a pseudo-random number generator 
(e.g.,~\cite{Marsaglia:2003:JSSOBK:v08i14})
seeded using the previous block's hash XOR'd with the acceptor signature of the
first selected transaction.
This scheme resists manipulation because no party can control the generator's seed.

Next, process the transactions in the permutation order,
deferring consideration of any transaction that cannot yet be processed
because its nonce is out of order.
Deferred transactions remain in permutation order.
Once all transactions have been considered,
the deferred transactions are similarly processed in
permutation order,
again deferring any transaction that still cannot
be processed because its nonce is out of order.
This process is repeated until all transactions have been processed,
or until a full pass over the deferred transactions yields no more processable transactions.
Since there is no way to order the remaining transactions,
they can be explicitly rejected.

In principle, this process could take $O(n^2)$ time for $n$ transactions,
but such behavior is highly unlikely.
It would require the selected and permuted transactions to include a
subsequence of $O(n)$ transactions,
each of which depends on its succcessor.
So many dependent transactions rarely happen,
and it is even more unlikely that a random permutation would produce such an ordering.

\idtonly{Persons skilled in the art will appreciate a variety of
  alternative mechanisms for randomizing transaction order and for
  restoring any dependencies violated by doing so.  For example,
  transactions could be sorted according to any function that nobody
  can control.  One example among many possibilities for randomizing
  the order would be to compute the difference between an
  appropriately chosen number of bits of a transaction's hash or
  acceptor signature and some value dependent on other factors beyond
  anyone's control, such as the previous block's hash.  To restore
  violated dependencies while avoiding the potential $O(n^2)$ worst
  case discussed above, we can observe that a transaction that spends
  from $k$ accounts enables at most $k$ more transactions (a
  transaction that spends from each of the input accounts using the
  nonce value following the one in the input).  This fact can be
  exploited to define asymptotically superior transaction ordering.
  For example, transactions could be inserted into a forest in which
  each $k$-input transaction has at most $k$ children, representing
  the dependencies discussed above, and transactions ordered by a
  deterministic traversal of all trees in the forest, with each
  transaction ordered before any of its children.}

%%% Local Variables: 
%%% mode: latex
%%% TeX-master: t
%%% End: 

\section{Discussion}
\label{sec:discussion}

\paragraph{Performance} We have not implemented our ideas
sufficiently to evaluate their impact on performance.  On one hand,
they eliminate the need for every transaction to be executed by every
node as it propagates through the network.  On the other hand, it also
adds overhead in various ways.  Decentralized ledger technology has
the potential to dramatically reduce cost and latency of transactions
by eliminating the need for trusted intermediaries that often add days
to transaction settlement times.  Thus, we should optimize for trust
and accountability before focusing on smaller performance
improvements\vsp.

\paragraph{Resolving Differences} We have focused on how to
make it difficult to manipulate outcomes without detection,
while enabling participants to \emph{explicitly} reject transactions,
and perhaps provide a reason, as may be needed to comply with regulations.
We have not addressed the question of
how a proposer should resolve a transaction's outcome if it receives
the transaction normally from some peers and rejected from others.
Another question is whether a transaction could be included by a
subsequent proposer after being rejected in an earlier block.

These questions are primarily ledger-specific policy matters, but they also
intersect with implementation details.
These issues illustrate the benefits of avoiding monolithic designs
by structuring systems so that different components can be instantiated
differently for different use cases.
Tendermint has moved in this direction after the version considered in this paper
\cite{Tendermint-TMSP}.
We expect that the ideas described here carry over,
although we have not yet explored this direction in any detail\vsp.

\paragraph{Acceptor Manipulation}
No amount of tamper-proofing will benefit a transaction that has not been
accepted by an acceptor.
An acceptor could refuse to accept or even acknowledge a transaction.
More subtly, it could attempt to manipulate acceptance order by
delaying transactions before assigning an acceptor ID.  However,
because acceptor order does not affect the final order of
transactions included in the same block (see
Section~\ref{sec:ordering}),
the acceptor's only reliable means of influencing transaction
order is by significantly delaying transactions.
It may be difficult to prove manipulation in individual cases,
but repeated patterns of manipulation would emerge from long-term analysis.

Front-running, where a node injects a transaction of its own
before another one, is difficult for acceptors because they have
little control over the eventual relative order of the two
transactions; front-running possibilities for proposers are
discussed next\vsp.

\paragraph{Proposer Manipulation}
A proposer can reorder transactions it has not yet confirmed.  It
could also insert transactions accepted by itself or an ally at the
end of its outgoing queue.  Thus, if the selection policy sometimes
allows a proposer to select (almost) its \emph{entire} outgoing queue,
then it has some latitude to trial different selections and choose
between them after determining the resulting order for each selection
(\secref{ordering}).

To mitigatate this possibility, we could further constrain selection.
A proposer could be required to exclude transactions accepted by
itself, or to order them last.  It could also be forbidden from
including unconfirmed transactions, or those for which confirmations
have not yet been acknowledged.  Alternatively or in addition,
policies and parameters could be chosen so that, in the common case,
selected transactions have already been included in a summary sent
to peers and acknowledged,
ensuring opportunities for manipulation such as front running
are rare and thus more
apparent when they do occur.

Permissioned acceptors and proposers suspected of manipulation can be held
accountable in various ways in addition to mechanisms that penalize them directly.  They may lose business or face
regulatory scrutiny or community pressure.  If a
substantial fraction of acceptors misbehave, trust in the overall
network will be diminished, reflecting that the choice of permissioned
participants was or has become poor\vsp.

\paragraph{Integrating External Decisions} Permissioned ledgers aim to
take advantage of existing trust and accountability mechanisms, while
recognizing that they can never be absolute.  We have focused
primarily on technical means for making undetectable violation of rules difficult.
In some cases, such violations
can be proved, allowing for ``transactional punishment''
without human intervention.  For example, Tendermint already allows a
node that signs conflicting votes to be automatically
penalized via an administrative transaction that confiscates its
``stake''\negthinspace; other provable infractions we have enabled can
similarly be automatically punished.  Nonetheless,
some violations, including acceptor manipulation, can be punished
only based on judgments made outside the system (informed
by evidence produced by the system).

This raises the question of what can be done when such external judgments
are made.  Again, this is a policy matter that
intersects with technical details.  For
example, the rules governing a given ledger might enable a majority of
participants to impose a penalty on one party, or, say five out of
seven members of a governance board to permanently exclude the party
that is judged to be behaving dishonestly.

No technical solution can prevent or solve all human disagreements,
but by making the ledger's rules precise,
and violations detectable,
many disagreements might be avoided in the first place,
and mechanisms that support effective governance
can facilitate resolution even when human intervention is needed\vsp.

\paragraph{Legitimate Losses} As described so far, transactions could
be lost through no malicious intent.  An acceptor might crash
immediately after sending a response and before sending the
transaction to its peers.  Should it be held responsible for the loss
of the transaction in this case and if so how?

If transaction loss is sufficiently undesirable, an acceptor may
be heavily penalized for such a loss, giving it an incentive to ensure
this does not happen.  It could do so by persisting the transaction
before sending the receipt, which would add significant overhead using
traditional storage technologies.  As nonvolatile RAM
\cite{nvmsurvey2014} becomes cheaper and more widely available,
avoiding such losses will impose less overhead, thus reducing the
level of penalties needed to encourage avoiding it\vsp.

\paragraph{Fault-Tolerance} As described, our changes make the
protocol susceptible to failures, such as a node crashing and missing
messages.  To resume normal operation, a restarting node can report
the outage via administrative transactions; downstream nodes may
forgive the omissions, which are made visible and accountable by
the administrative messages.  If a node omits a transaction
without explanation,
its peer may insert an
administrative transaction to report the omission and continue normal
operation.  Either way, evidence is created to shed light on the reason
for the omission.  

%%% Local Variables: 
%%% mode: latex
%%% TeX-master: t
%%% End: 

\section{Related work}
\label{sec:related}

To our knowledge, the blockchain-related work most closely related to ours is Factom
\cite{factom-whitepaper-1.0}, a ``proof of publication'' system that
aggregates ``entries'' submitted to ``federated servers'' and records
hashes in the Bitcoin ledger to support undeniable proof of the
entries' existence at a certain time.  Factom's approach to preventing
manipulation is similar in spirit to ours.  For example, Factom
servers produce ``confirmations'' of entries that are ordered, and
ensures that servers cannot ``lose'' them or manipulate their order
without detection.  It also incorporates mechanisms for proving
violations of the rules and penalizing the culprits automatically, as
well as for allowing the ``community'' to remove a server if it loses
support for any reason.

Factom allows clients to buy ``entry credits'' (using tokens called
``factoids''), which reserve the right to submit an entry,
without revealing identity or content.  Instead, a
hash of the entry is included, allowing clients to reveal the original
content after the entry has already been included in the ledger.

Similar techniques could be used with our system to acquire
acknowledgment of an intended transaction from an acceptor before
revealing the transaction.  However, a key difference for our context is that
the semantics of transactions and smart contracts depend on ordering; thus
to guarantee ordering to a transaction without
being able to process it requires a fundamentally different approach.  

It might seem that projects such as Enigma \cite{Enigma} and Hawk
\cite{cryptoeprint:2015:675}---which entail techniques for processing
transactions without knowing their contents---might provide an
alternative way to achieve similar accountability benefits while
supporting transactions.  However, such approaches do not preclude the
possibility of validators favoring transactions of their allies,
because they can identify the transactions with help from their
allies, even though the transactions may be encrypted (in part).

Techniques used by our OWAC protocol---as well as for representing
nodes' outgoing queues---are reminiscent of previous work (e.g.,
\cite{Lamport:1982:BGP:357172.357176,Merkle:1987:DSB:646752.704751,Martel:2004:GMA:997201.997203,Miller:2014:ADS:2535838.2535851,Anagnostopoulos:2001:PAD:648025.744371})
in which new data includes hashes of previous data, thus preventing
tampering, and cryptographic signatures are used authenticate data,
thus preventing forgery.

We have explored the use of such techniques specifically for
enhancing accountability in distributed ledgers.  Previous work has
addressed accountability in distributed systems more generally.
Yumerefendi and Chase \cite{Yumerefendi:2005:RAD:1973400.1973403} propose that accountability be considered a
``first-class design principle'' for dependable network systems.  The
design principles espoused in this paper are consistent with that
vision.

Reiter and Birman \cite{Reiter:1994:HSR} identify attacks that can
cause honest servers to misbehave by violating causality when
delivering messages, citing front-running as an example.  They
propose a solution based on threshold cryptography, whereby honest
servers participate in decrypting messages only after committing to
deliver them.   
% (using a scheme due to Lim and Lee [Crypto '93] to avoid
% chosen-ciphertext attacks, whereby dishonest server learns shares from
% others by seeing how they contribute to decryption of a message of its
% choosing).  
Cachin et al. \cite{DBLP:conf/crypto/CachinKPS01} present efficiency
enhancements and more formal analysis.  These approaches assume an
underlying atomic broadcast protocol, and thus inherit their
assumptions and overhead in addition to their guarantees.  Our
approach is lighterweight and more flexible: we do not attempt to
ensure all transactions are delivered to all participants in the same
order.  Instead, we impose rules for ensuring that the orders in which
transactions are propagated and included in blocks are not
manipulated, and propose mechanisms that discourage accountable
entities from violating these rules by detecting (potential)
violations and holding perpertrators accountable.

Haeberlen et
al. \cite{haeberlen-2006-detection,haeberlen-2007-peerreview}
similarly aim for lightweight approaches that focus on detecting
misbehavior in ways that perpertrators can be held accountable, rather
than on preventing or masking it.  Our work is similar to theirs in
that participants are required to maintain authenticated, tamper-proof
records that can be examined by others to detect misbehavior.  Some of
the techniques we propose are reminiscent of optimizations used by
Haeberlen et al., including buffering logs before sending, and
allowing participants to prove claims about data without exchanging
entire data structures.

Our work also differs in several ways.  Haeberlen et al. propose
protocols that aim to always eventually detect a byzantine participant
while never falsely accusing an honest one, addressing questions such
as when participants should challenge each other and how they should
respond.  In contrast, we have described mechanisms for proving
misbehavior in some cases and exposing evidence that may be
accummulated and interpreted externally in others, while not
addressing how this should be done.

Haeberlen et al. propose a general methodology in which correct
behavior of each participant is defined by a deterministic state
machine, and is held accountable for following its transitions
faithfully according to messages received.  In contrast, our work
specifically addresses integration into a permissioned distributed
ledger context, exemplified by Tendermint.

Neither our ideas nor those of Haeberlen et al. can prevent a
participant from manipulating the order in which concurrent messages
are received from different peers.  However, our work includes
mechanisms that severely limit the ability of participants to exploit
such manipulation.  In particular, we ensure that validators cannot
control the ordering of transactions within a block even if they
manipulate the order in which they claim to receive them.

We also introduce the idea of accountable censorship, which is an
application-level issue that is not addressed by a generic approach to
state machine logging for an existing application.

%%% Local Variables: 
%%% mode: latex
%%% TeX-master: t
%%% End: 

\section{Concluding remarks}
\label{sec:concluding-remarks}
Designing distributed ledgers to be resistant to internal fairness violations is a complex and important problem,
and this paper is only a first step.
We exploit the common-sense observation that permissioned ledgers,
which have recourse to real-world measures such as fines, expulsions, or legal action,
can rely on unobtrusive, after-the-fact auditing and statistical analysis to detect and deter misbehavior,
especially repeated misbehavior.
By reducing the number of non-deterministic choices,
auditors have fewer possibilities to consider when anomalies are detected.
To this end,
we have proposed several deterministic mechanisms intended to make violations more cumbersome and easier to detect.
There are many directions for future work.
Can we devise more rigorous, more lightweight mechanisms?
What other kinds of fairness violations need to be deterred?
Are there theoretical bounds on the kinds of fairness properties that can be monitored,
and the cost of such monitoring?
%%% Local Variables: 
%%% mode: latex
%%% TeX-master: t
%%% End: 

% References stuff
%\newcommand{\bibfont}{\scriptsize}
%\setlength{\bibsep}{2pt}

\bibliographystyle{plain}
\bibliography{blockchain-censorship}

\end{document}